\documentclass[letterpaper]{article} %
\usepackage{aaai24}  %
\usepackage{times}  %
\usepackage{helvet}  %
\usepackage{courier}  %
\usepackage[hyphens]{url}  %
\usepackage{graphicx} %
\usepackage{natbib}  %
\usepackage{caption} %
\usepackage{algorithm}
\usepackage{algorithmic}

\newcommand{\descr}[1]{\smallskip\noindent\textbf{#1}}
\usepackage{xcolor}
\usepackage{booktabs}
\usepackage{cleveref}

\newcommand{\answerYes}[1]{\textcolor{blue}{#1}} 
\newcommand{\answerNo}[1]{\textcolor{teal}{#1}} 
\newcommand{\answerNA}[1]{\textcolor{gray}{#1}} 

\usepackage{newfloat}
\usepackage{listings}
\DeclareCaptionStyle{ruled}{labelfont=normalfont,labelsep=colon,strut=off} %
\floatstyle{ruled}
\newfloat{listing}{tb}{lst}{}
\floatname{listing}{Listing}
\title{Exploring Climate Change Discourse: Measurements and Analysis of Reddit Data}
\author{
    Smriti Janaswamy,
    Jeremy Blackburn
}
\affiliations{
    Binghamton University\\
    sjanasw1@binghamton.edu, jblackbu@binghamton.edu
}

\begin{document}

\maketitle

\begin{abstract}
Social media is very popular for facilitating conversations about important topics and bringing forth insights and issues related to these topics. 
Reddit serves as a platform that fosters social interactions and hosts engaging discussions on a wide array of topics, thus forming narratives around these topics.
One such topic is climate change. 
There are extensive discussions on Reddit about climate change, indicating high interest in its various aspects. 
In this paper, we explore 11 subreddits that discuss climate change for the duration of 2014 to 2022 and conduct a data-driven analysis of the posts on these subreddits. 
We present a basic characterization of the data and show the distribution of the posts and authors across our dataset for all years.
Additionally, we analyze user engagement metrics like scores for the posts and how they change over time. 
We also offer insights into the topics of discussion across the subreddits, followed by entities referenced throughout the dataset. 
\end{abstract}

\section{Introduction}
Climate change refers to the long-term changes in global and regional climate patterns, primarily due to human activities, also referred to as ``anthropogenic climate change''~\cite{milfontPublicBeliefClimate2017, capstick2015international, incAwarenessClimateChange2009}. 
It is considered one of the biggest threats to our well-being be it environmental, economic, or political~\cite{fernandezTalkingClimateChange2016, ferrari-lagosEducationMobilizeSociety2019}. 
The Intergovernmental Panel on Climate Change (IPCC) considers the 1--degree increase in global warming above pre-industrial levels to be primarily due to human activities~\cite{Summaryf7:online, IPCCIntergovernmentalPanel}, thereby indicating that climate change is no longer an issue to be stopped, but something to be mitigated. 
Impacts of climate change range from a rise in temperatures and sea level, changes in rainfall and frequency of other weather-related events like storms, droughts, or floods to social or economic impacts like aggravating the effect it has on vulnerable and low--income populations, and marginalized communities~\cite{rahmstorfKeynoteClimateChange2012, Whatthew35:online}. 
As global temperatures rise and extremes in weather-related events become more frequent, comes the need for understanding climate change to effectively mitigate and adapt to climate change. 

There are many important topics on social media about climate change, which can be widely found on many platforms where discussions on climate change and its various aspects take place. 
Climate related events like the Paris Climate Agreement and Climategate are well documented and are of substantial importance in the research community~\cite{koteykoClimateChangeClimategate2013, hopkeParisClimateTalks2017, leiserowitzClimategatePublicOpinion2013, treen2022discussion, niedererGlobalWarmingNot2013}. 
Despite the impact of these events, there are not many studies that examine these climate related discussions on social media in detail. 
To address this gap, we study climate change discussions on Reddit. 
We analyze 11 subreddits that talk about climate change for the period 2014 to 2022. 
Of the 11, we start with a seed list of 5 and use empirical methods to find more that are relevant to climate change and talk about various aspects of it. 
We do quantitative topic analysis using BERTopic, a topic modeling technique. 
Given the large size of our dataset, looking at topics obtained as a result of applying this technique helps us go through multiple topics via keywords, organize, and understand them at scale. 
We perform this analysis by the year and compare those that best represent the topics from a specific year. 
We compile the top 10 topics by year and observe that some topics like wildfires, the Paris Agreement, ice melting, solar panels, renewables, etc. repeat in consecutive years. 
We also see that Europe is a highly cited location across various topics.
Our results indicate the significance of these issues and emphasize the necessity of acknowledging or addressing these concerns. 
\section{Background and Related Work}
In this section, we address previous research on climate change and related activities. 
First, we talk about climate change and research on its various aspects. Subsequently, we discuss the role social media plays in climate change discourse before examining the relevant works in this domain. 

\descr{Climate Change.} 
Ongoing research efforts cover the various causes and consequences of climate change, and identify strategies for mitigation. 
\cite{dunlapPoliticalDivideClimate2016, Whatthew35:online, tyagiAffectivePolarizationOnline2020} show correlations between partisan divides and Americans' beliefs regarding climate change polarization.
They state that the increasing political polarization brings about more divide in opinions on anthropogenic climate change, with Democrats seeing increasing numbers of those believing in it, and a decline among Republicans who believe this to be true. 
They also show that Democrats believed more than Republicans that global warming had officially begun and that changes in the Earth's temperatures were attributed to human activities. 
\cite{treen2022discussion} study the nature of the relationship between climate change and political polarization resulting from skepticism about climate change, finding that these two are linked to misinformation. 
Skepticism exists among people who believe any policies in place for climate action will deter them from personal freedom~\cite{hoffman2011culture}. 

Activism for climate change refers to the efforts undertaken by people - either individuals or as a group or organization to increase awareness about climate change and its impacts. 
This entails advocating for policy changes that will help mitigate the impacts. 
Other ways activism is pursued is through educating people about the science behind climate change and the need for urgency in taking action to address it~\cite{hestres2017internet, stoddart2016canadian, walter2018echo}. 

\descr{Research on Climate Change and Social Media}
Social media platforms like Facebook, Twitter, Instagram, and Reddit have a large number of users spending a lot of time engaging and relying on them as their primary source of news consumption.
Users leverage these platforms to exchange opinions and raise awareness about important events~\cite{mavrodievaRoleSocialMedia2019}. 
This gives us an incentive to study climate change on social media.
Prior research has been conducted on various aspects of climate change, primarily on Twitter data~\cite{cannIdeologicalBiasesSocial2021, dahal2019topic, pearce2014climate, cody2015climate, walter2019scientific, fownes2018twitter}. 
\cite{cody2015climate} study climate change on Twitter, find it to be a topic highly politicized in the United States and often associated with an attitude of denial towards climate change. 
They examined tweets that contain the word ``climate'' often associated with the word ``denial'' or ``deny.'' 
\cite{stede2021climate} identify discourses on climate change to be one of: 1) media, 2) individuals or social communities engaging in discussions, and 3) discourse identified by the stance on the topic (believe or deny). 
\cite{tuitjerSocialMediaPerceived2021} examines the impact of posts on social media platforms like YouTube, Facebook and Instagram on the perception of climate change across Europe. 
Apart from news and other avenues for media coverage, Twitter has been the platform of choice for research on climate change and has been used in numerous studies. 
\cite{dahal2019topic} study climate change on Twitter and take locations from tweets into consideration to perform text mining techniques like sentiment analysis and topic modeling. 
They found that climate change related discussions on Twitter are negative in nature and topics of discussion covered many aspects like the environment, weather, and beliefs about the topic. 
\cite{roxburghCharacterisingClimateChange2019} also study climate change on Twitter to learn more about weather related events like Hurricane Irene, Hurricane Sandy, Snowstorm Jonas and climate change. 
They note the increase in Twitter's user base after a major weather related disaster. 
\cite{yla2022topic} study media coverage in India and USA to study various topics of discussion via frame analyses.
\cite{uthirapathy2023topic} perform topic modeling using LDA and sentiment analysis to uncover popular topics and gain insights into the sentiments across tweets pertaining to climate change. 
There are studies that have worked on understanding climate change using Reddit as a data source~\cite{treen2022discussion, parsaAnalyzingClimateChange2022} to gain an understanding of how climate change is perceived in the discussions. 
These have also used LDA for topic modeling. 
\cite{jiang2017comparing} analyze various topics pertaining to climate change using data from newspaper articles and performs sentiment analysis on this data across the various newspapers included. 
\cite{bohr2020reporting} also conducts an analysis of topics related to climate change using news articles and find differences in the coverage of news pertaining to aspects of climate change. 
In addition to Twitter, news sources, there are studies conducted on climate data from Facebook as well. 
\cite{islamAnalysisClimateCampaigns2023} look into Facebook ads as a medium to disseminate information about climate change. 
One example of activism for climate change was the \emph{\#FridaysForFuture} movement started by Greta Thunberg which inspired people across the world to protest for political action to address climate change challenges~\cite{haugestad2021youth}. 
\cite{spence2012psychological} study how people think about climate change by looking at when a climate change related event will occur, where (geographically) it will occur, and how uncertain people are about it. 
They find that how people decide to take action about a climate change event is directly related to when an event takes place. 
For example, a weather-related event like drought or flood can initiate a sense of urgency in people and have them wanting to do something about it. 
They also find that educating people by helping them understand how climate change can affect them directly can motivate them to take action. 
In the midst of growing concerns about climate change, social media like Reddit has proven to be a valuable source of data for studying public perceptions, stance and attitudes, and discussions related to the environment and climate change. 

\section{Data}
We select Reddit as our data source because of its popularity, the vast user base and the diverse range of content across various communities known as subreddits. 
There are over 130,000 subreddits~\cite{balci2023data} that organized based on the topics of discussion that can take place and are denoted by \emph{r/}~\cite{baumgartner2020pushshift}. 
As of December 2023, Reddit was one of the most visited websites with approximately 2.1 billion visits~\cite{Redditco41:online}. 
Data on subreddits has conversations via ``posts'' and ``comments'' on said posts and users can upvote or downvote posts. 
Each subreddit has its own set of rules to monitor the nature of content one can post. 

\subsection{Seed Subreddits}
For our study we start with a set of subreddits that serves as a seed list: 1) r/climate, 2) r/environment, 3) r/climate\_science, 4) r/climatepolicy and 5) r/climateskeptics. 
We choose these subreddits for our seed list  because they provide not only climate-specific information but also insights pertaining to the environment as a whole, reflecting on the multidisciplinary nature of climate change. 
This means that climate change as a topic requires integration of knowledge and perspectives from various fields of study like Earth sciences, environmental science, ecology, policy, governance and atmospheric science.
Of the ones in our seed list, the one that first came into existence is r/environment in 2008, followed by r/climate. 
Of these, r/climate has previously been used in research~\cite{treen2022discussion} and has posts primarily from a global standpoint, including topics like activism and climate--related politics. 
r/climate was created on May 2008 and has around 187K members. 
The sidebar on the subreddit page lists this as part of the top 1 percent of all subreddits. 
The second subreddit we choose is r/environment, which has posts that have information about the environment with topics ranging from temperature changes, wildfires, climate action to climate activism, air pollution, and plastic pollution, etc.
It dates back to Jan 2008 and has around 1.6M members. 
This subreddit is also part of the top 1 percent of all subreddits. 
Subreddit r/climatepolicy is a smaller subreddit that was created on September 2015 and has around 1.6k members. 
Posts on this subreddit mostly concern policies surrounding climate change and their effectiveness. 
Subreddit r/climateskeptics is for those who question climate change and had its first post in July 2008. 
Finally, there is a subreddit for climate change science and news called r/climate\_science which is the last one in our seed list.

\subsection{Finding Related Subreddits}
While our current list of subreddits is good in that it provides us with an overview of what kind of posts and discussions they entail, we recognize that there may be more subreddits that will help enhance our understanding by providing more information and cover some more aspects of climate change. 
To this end, finding subreddits that discuss climate change indirectly can be challenging as they may not be explicitly related to climate change, but may touch upon its various aspects or related topics. 
Users that post on the seed communities may also post in other communities about an aspect of climate change, and they may not be explicitly related to climate change either. 
We aim to find such communities to analyze for our study and do so by making use of \emph{subreddit embeddings} to better understand similarities between various subreddits, detect new communities and understand behavior of users that post on similar subreddits~\cite{balciFishBicyclesExploring2023a, waller2021quantifying}. 
The idea behind this is that users posting on one subreddit implies interest in that topic and may post on related subreddits too. 

\subsubsection{Subreddit Embeddings}
We take inspiration from \cite{balciFishBicyclesExploring2023a} and make use of pretrained embeddings provided by \cite{raymondMeasuringAlignmentOnline2022b}, which uses the same techniques as Waller and Ashton~\cite{waller2021quantifying} for training these embeddings. 
These embeddings use users as the context and subreddits as words~\cite{levy2014dependency}. 
We take this approach because all the subreddits in our seed list were created by this time, are actively used, and have some substantial amount of data to uncover patterns from comments posted on the subreddits. 
This factor is important to our study because we collect data from 2014.
The 10,000 subreddits that are used to create this embedding contain three out of five of our seed subreddits: 1) r/climate, 2) r/environment, and 3) r/climateskeptics.

\subsubsection{Clustering}
We follow methodology by \cite{balciFishBicyclesExploring2023a} and apply dimensionality reduction with UMAP~\cite{mcinnes2018umap} followed by agglomerative clustering, a hierarchical form of clustering that works in a bottom up approach by starting with treating each data point as a cluster and merging close clusters at each step until all the data points are covered. 
We attempt to cluster incrementally by starting with 10 as the desired number of clusters and keep incrementing this by 10 until we get the largest number of clusters for which all of our seed subreddits end up in the same cluster. 
This results in 130 clusters.
The cluster containing all the seeds includes 218 subreddits. 
We manually inspect the 218 subreddits and find that there are some more subreddits suitable to our study because they talk about climate change topics like r/climatechange, r/ClimateActionPlan, and r/ClimateOffensive. 
In addition, we find subreddits that talk about aspects of climate change like sustainability and renewable energy. 
We add r/sustainability, r/energy, and r/RenewableEnergy to our list as well, leaving us with 6 additional subreddits.

\subsection{Data Collection}
We collect posts from 2014 to 2022 for 11 subreddits using the Pushshift API~\cite{baumgartner2020pushshift}. 
Table~\ref{tab:sub_overview} shows the details of the subreddits with timelines we include in our collection. 
Details like the number of posts and the number of authors that make posts in each of the subreddits are also listed. 
The subreddit r/environment has the most number of posts and authors that post in our dataset, followed by r/energy. 
The least number of posts are in r/climatepolicy and r/climate\_science. We include them in our dataset inspite of this low number.

\begin{table}[t]
    \begin{tabular}{llll}
    \toprule
    Subreddit & \# Posts & \# Users & Years \\
    \midrule
    r/climate & 64,100 & 11,190 & 2014 -- 2022 \\
    r/environment & 206,559 & 38,411 & 2014 -- 2022 \\
    r/climate\_science & 2,505 & 1,271 & 2015 -- 2022 \\
    r/climatepolicy & 1,623 & 382 & 2015 -- 2022 \\
    r/climateskeptics & 43,362 & 4,546 & 2014 -- 2022 \\
    r/climatechange & 22,749 & 10,154 & 2014 -- 2022 \\
    r/ClimateOffensive & 9,257 & 3,071 & 2018 -- 2022 \\
    r/ClimateActionPlan & 10,407 & 3,722 & 2018 -- 2022 \\
    r/energy & 81,323 & 14,544 & 2014 -- 2022 \\
    r/RenewableEnergy & 26,685 & 7,468 & 2014 -- 2022 \\
    r/sustainability & 23,787 & 11,417 & 2014 -- 2022 \\
    \bottomrule
\end{tabular}
\caption{Details of the subreddits in our list as they appear in our dataset. The Years column indicates how many years of data we include for each. We collect data starting 2014 and until 2022. Subreddits r/climate\_science and r/climatepolicy do not start from 2014 and so, we have collected them from their creation date. The same goes for r/ClimateOffensive and r/ClimateActionPlan which were created in 2018. The Users column indicates the number of authors that post on each of the subreddits over the years. The highest number of authors is seen for r/environment and the lowest for r/climatepolicy with a count of 382.}
\label{tab:sub_overview}
\end{table}

\section{Data Analysis}
In this section we provide an exploratory analysis of our subreddits and the content posted to them. 

\subsection{Basic Characterization of our Data}
\begin{figure*}[t]
    \centering
    \includegraphics[width=2.3\columnwidth, bb=0 0 1830 480]{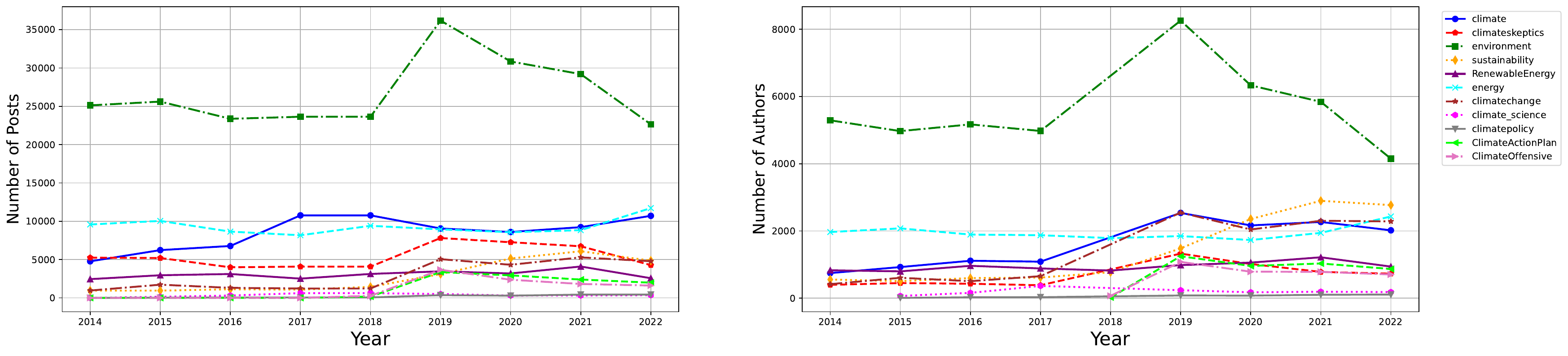} 
    \caption{Number of posts and authors across all subreddits from 2014 -- 2022.}
    \label{fig:posts_authors_plot}
\end{figure*}

\begin{figure}[t]
    \centering
    \includegraphics[width=\columnwidth, bb=0 0 730 390]{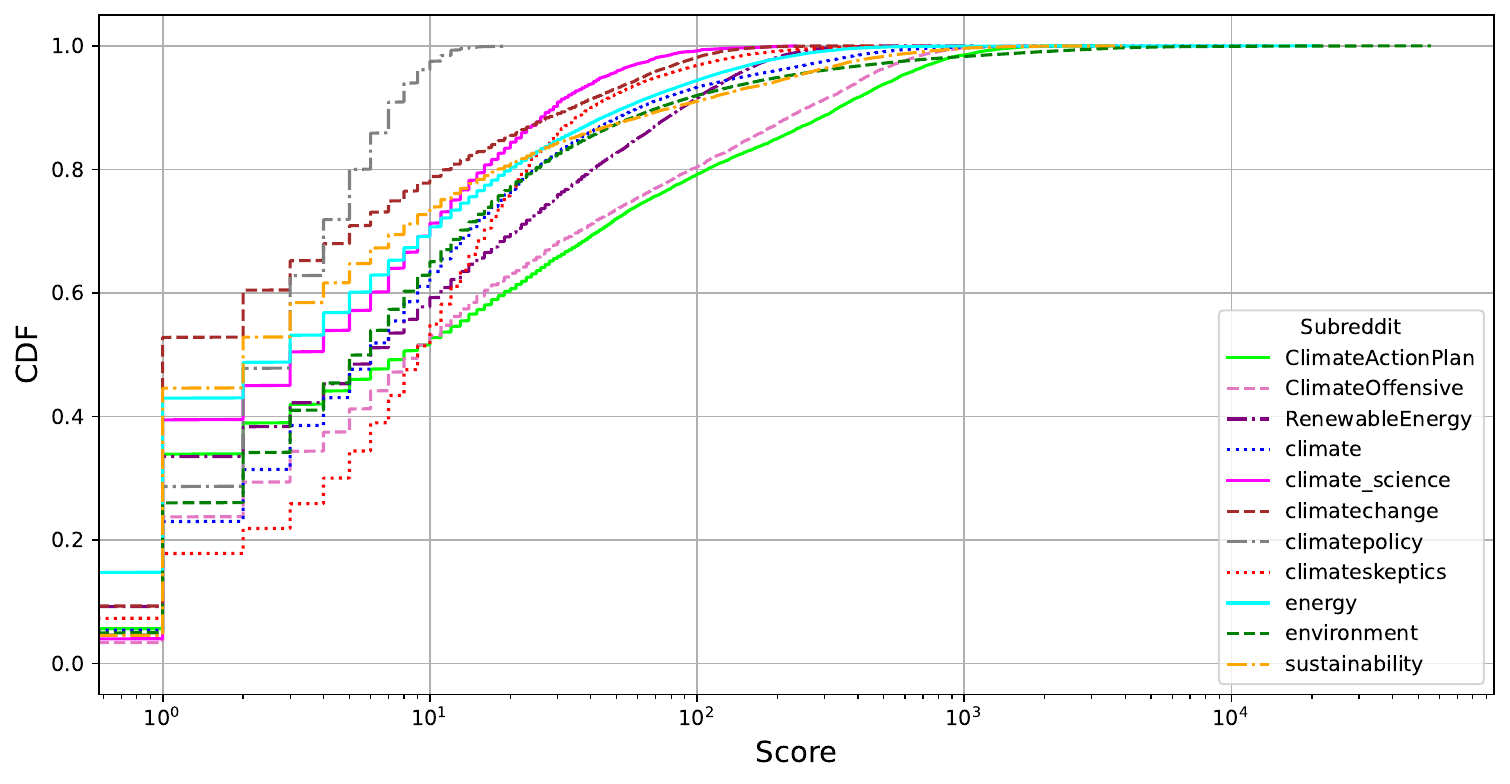} 
    \caption{CDF of submission scores for all subreddits.}
    \label{fig:subr_cdf}
\end{figure}

\begin{figure}[t!]
    \centering
    \includegraphics[width=\columnwidth, bb=0 0 840 620]{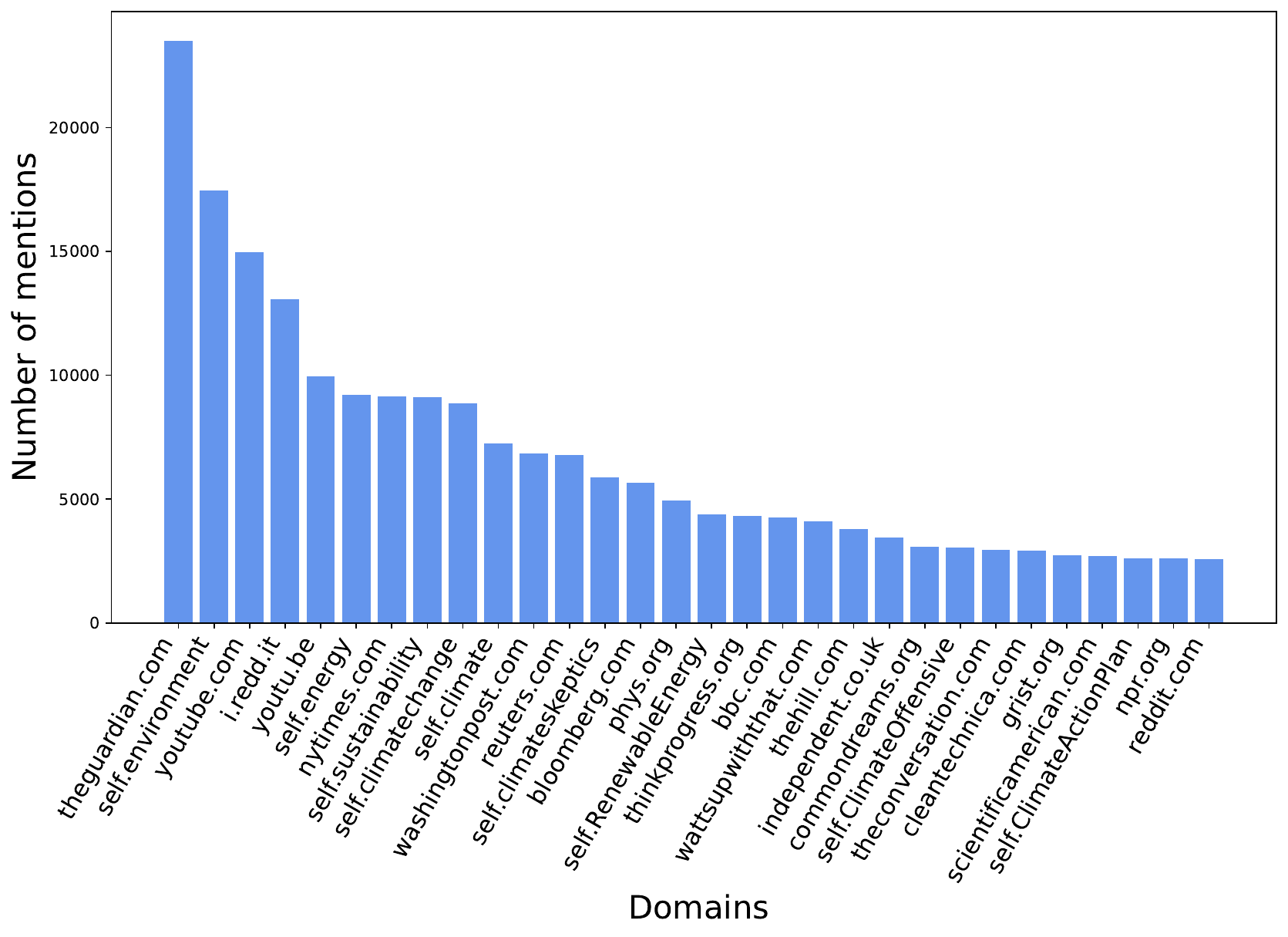} 
    \caption{Top domains referenced in posts in our dataset. Shown here in descending order.}
    \label{fig:domains30}
\end{figure}

Figure~\ref{fig:posts_authors_plot} shows the number of posts per year and the number of authors per year for each of the subreddits. 
We note that r/environment consistently has the highest number of posts in comparison to the others. 
Subreddits r/climatepolicy and r/climate\_science have fewer posts in comparison and were also created much later, in 2015. 
Note that r/environment has the most authors posting.  
Figure~\ref{fig:subr_cdf} shows the cumulative distribution function (CDF) of submission scores for each of our 11 subreddits.
Figure~\ref{fig:domains30} presents a list of the top 30 domains that are referenced in the posts from our dataset. 
The most referenced domain is \emph{theguardian.com} with a count of 23,506. 
By \emph{Self.environment}, we mean the subreddit r/environment itself and it is referenced 17,463 times.
\emph{Youtube.com} is third on the list and is referenced 14,962 times. 
Others referenced in the list of top 30 include \emph{wattsupwiththat.com}\cite{WattsUpW32:online}, \emph{thehill.com}, \emph{independent.co.uk}, \emph{commondreams.org}, \emph{thinkprogress.org}, etc. 
\emph{wattsupwiththat.com} primarily features information related to skepticism about climate change and often has content that challenge mainstream climate science. 
The website covers a range of topics, including weather patterns, climate sensitivity, the Climategate controversy, etc.
Content on the main page can be filtered using some popular hashtags provided like \emph{\#climate change}, \emph{\#Global warming}, \emph{\#Intergovernmental Panel on Climate Change}, \emph{\#NASA}, etc.
The Climategate controversy unfolded in 2009 before the Copenhagen Summit, and involved the unauthorized release of emails spanning several years from climate scientists associated with the University of East Anglia's Climate Research Unit in the UK. 
These emails were illicitly obtained and subsequently made public, fueling debates and furthering climate denial agendas~\cite{heronThingsFallApart2023, leiserowitzClimategatePublicOpinion2013, TheImpac46:online}. 
\emph{commondreams.org} is a U.S. based website and a non--profit that provides progressive perspectives on various issues like politics, social justice, the environment and climate change. 
Similarly, \emph{thinkprogress.org} was founded in 2005 and ran until 2019 that discussed politics and has a section on climate--related news called ``Climate Progress''~\cite{ThinkPro62:online, ThinkPro99:online}.

\subsection{Topic Analysis}
We analyze the data from the subreddits to uncover prevalent topics in the discussions. 
To achieve this, we train a BERTopic~\cite{grootendorst2022bertopic} model that helps find topics from the documents using the following steps: 1) extracting the document embeddings, 2) apply dimensionality reduction using UMAP, 3) cluster the reduced embeddings using a density based clustering algorithm called HDBSCAN~\cite{mcinnes2017hdbscan} and then, 4) extracts topics from the cluster. 
We use default parameters while training the model. 
We use the titles of all the posts from our subreddits to find topics and remove posts generated by AutoModerators as they don't contribute much to the nature of posts we wish to analyze for topics. 
Posts marked as ``[deleted]'' are retained in our analysis because even though the post may be deleted, the titles persist, and we make use of these titles as well in our analysis.

Table~\ref{tab:topics_across_years} lists the total number of topics across our data by the year. 
The earlier years do not generate as many topics as the later years. We see the most topics generated for the year 2019 for a total of 1,282. 
The least number of topics are generated for the year 2014.
The count reflected here does not include the first set of topics returned after running the model. 
This is because it is outliers and is referred to as ``Topic -1.''

\begin{table}[t]
  \centering
  \begin{tabular}{cc|cc|cc}
  \toprule
  \textbf{Year} & \textbf{\# Topics} & \textbf{Year} & \textbf{\# Topics} & \textbf{Year} & \textbf{\# Topics} \\
  \midrule
  2014 & 723 & 2017 & 798 & 2020 & 1,255 \\
  2015 & 746 & 2018 & 791 & 2021 & 1,175 \\
  2016 & 757 & 2019 & 1,282 & 2022 & 1,015\\
  \bottomrule
\end{tabular}
\caption{Number of topics generated for each year in our data.}
\label{tab:topics_across_years}
\end{table}

With BERTopic, each of the resulting topics has many words associated with it that represent the topic or come under the umbrella that is the topic name. 
By observing the keywords associated with each of the generated topics, we can infer topics of discussion in the subreddits. 
The representative docs that are provided with the keywords help understand how a topic is being talked about. 
\Cref{tab:topics_in_2014,tab:topics_in_2015,tab:topics_in_2016,tab:topics_in_2017,tab:topics_in_2018,tab:topics_in_2019,tab:topics_in_2020,tab:topics_in_2021,tab:topics_in_2022} in the Appendix show us the top 10 topics in the years 2014 to 2022. 
Each year's analysis consists of all the subreddits from our dataset, provided they have been created by that year. 
Table~\ref{tab:keywords_by_year} summarizes the top 10 topics by listing the topic labels for each of the years 2014 to 2022. 
We notice that topics like fracking (associated with Texas) drought, nuclear plants, melting ice, solar panels, etc. are popular in 2014. 
2015 sees some of these topics again like fracking (associated with Pennsylvania), carbon emissions, temperatures in July referring to the extremes in temperatures, the Paris Summit, and more. 
The Paris Summit, also referred to as the Paris Climate Agreement is a recurring topic of discussions across the years in 2015, 2016, 2017, and 2018. 
This agreement came into effect in December 2015 at the 21st U.N.'s Climate Conference. 
The shared goal was to combat and reduce global warming by limiting the increase in global temperatures to below 2 degrees Celcius. 
Other topics we see repeatedly over the years are electric cars, melting of ice, carbon dioxide levels, air pollution, etc. 
The evolution of topics reflects the changing narratives within climate change discussions. 
\begin{table*}[t]
    \centering
    \begin{tabular}{cp{14cm}}
    \toprule
    Year & Top 10 topics \\
    \midrule
    2014 & Fracking, Coal Spills, Oil Prices, Drought, Wind Turbines, Nuclear Plant, Solar Panels, Ice Melt, Renewables, Forests\\
    2015 & Carbon Emissions, Climate Skeptics, Fracking, Catholic Church, Coal, Warmest Year, Keystone Pipeline, Barack Obama, Paris Summit, Oil Prices \\
    2016 & Ice Melting, Batteries, Paris Agreement, Oil Prices, Donald Trump, Flooding, Veganism, Coral Reefs, Electric Cars, Sustainability \\
    2017 & Keystone Pipeline, Ice Melting, Air Pollution, Wildfires, EPA, Donald Trump, Batteries, Electric Vehicles, Paris Agreement, Nuclear Power \\
    2018 & Keystone Pipeline, Wind Turbines, Paris Agreement, Hydrogen, Electric Vehicles, Batteries, Donald Trump, Air Pollution, Wildfires, Contaminated Water \\
    2019 & Hydrogen, Renewable Energy, Carbon dioxide, Greta Thunberg, Nuclear Power, Green New Deal, Sustainable Clothing, Fewer Children, Greta Thunberg, Ice Melting \\
    2020 & Wind Turbines, Renewables, Oil Prices, Electric Vehicles, Carbon dioxide Levels, Climate Action, Green, Joe Biden, Sustainable Clothing, Fossil Fuel \\
    2021 & Wealthiest and Finance, Wind Turbines, Nuclear, Covid19 effects, Coal Mining, Electric Vehicles, Air Pollution, Sustainable Clothing, Greta Thunberg, Wildfires \\
    2022 & Wind Turbines, Forests, Sustainability, Renewables, Amazon Rainforests, Agriculture, Dubai and Climate Change, Wildfires, Electric Vehicles, Joe Biden \\
    \bottomrule
\end{tabular}
\caption{Summary of the top 10 topics' labels from 2014 to 2022.}
\label{tab:keywords_by_year}
\end{table*}

Despite temporal changes, we observe that certain topics remained persistent throughout the years. 
This can be seen as an indication of these topics being of paramount importance and emphasizes the need to highlight them. 
For e.g., the topic \emph{wildfires} appears in 2017, 2018, 2021, 2022 and in relation to California and drought.
This is associated with topics that mention \emph{forests} or \emph{deforestation}. 
Some representative documents for the topic \emph{wildfires} are:
\newline
\newline
\emph{Scientists See Climate Change in California's Wildfires}
\newline
\newline
\emph{People Cause Most U.S. Wildfires}
\newline
\newline
\emph{New Wildfires In Western U.S. At 10-Year High}
\newline
\newline
\emph{In Washington State, The 2014 Wildfire Season Has Been 6 Times Worse Than Normal.}
\newline
\newline
\emph{What's behind Europe's spate of deadly wildfires? Higher temperatures plus fuel accumulation}
\newline
\newline
\emph{LA mayor on Trump's response to wildfires: `This is climate change' not just about forest management}
\newline
\newline
\emph{It is just too much: Lucifer heat wave stifles parts of Europe. Extreme heat has fuelled wildfires, damaged crops, and strained energy and water supplies}
\newline
\newline
There are mentions of \emph{forests} in 2014, 2017, 2018, 2021, and 2022.
Representative docs pertaining to \emph{forests} include:
\newline
\newline
\emph{Save Trees. Save Earth.}
\newline
\newline
\emph{How much does planting trees really help?}
\newline
\newline
The keyword \emph{forests} is associated with other related terms like \emph{deforestation, reforestation, tropical, planting, trees}, etc.
The wildfires in California in 2021~\cite{2021Cali32:online} can be attributed to this topic showing up as one of the top during 2021 and 2022. 
They received 287 and 269 mentions, respectively. 
This topic is also mentioned in representative documents alongside mentions of Europe, Washington State, and Western U.S. 
Notably, we see that wildfires is said to occur as a result of anthropogenic climate change. 
It is also talked about as having more intensity than in previous years and is attributed to the higher temperatures.

Similarly, we notice this pattern of topics rolling into the next years with \emph{renewables} or \emph{renewable energy} a part of 2014, 2016, 2019, and 2020. 
We observe posts in the \emph{renewables} topic are generally along the lines of places committing to 100 \% renewable energy in the near future like Salt Lake City, San Diego, Beijing, India. 
Years 2016, 2017, 2018, 2020, and 2021 see \emph{electric vehicles} among the top 10 highly discussed topics across the subreddits. 
Mention of regions like the Arctic and Antarctic are talked about in relation to ice melting likely as a result of rising temperatures. 
Sea ice melting and rising levels of the water are also talked about in relation to these two locations, attributing them to climate change. 
Some sample representative documents that depict this are:
\newline
\newline
\emph{Antarctic sea ice at record levels despite global warming?}
\newline
\newline
\emph{Enough ice melted in Greenland on Tuesday to cover Florida in two inches of water, scientists warn | Climate News}
\newline
\newline
\emph{``Climate crisis: 2020 was joint hottest year ever recorded | Global heating continued unabated despite Covid lockdowns, with record Arctic wildfires and Atlantic tropical storms''}
\newline
\newline
Other topics related to extremes in temperatures are present as:
\newline
\newline
\emph{Are Tornadoes Changing Along With Climate? | Scientists have been able to draw links between a warming planet and hurricanes, heat waves and droughts, but the same can't be said for tornadoes yet.}
\newline
\newline
\emph{The strongest, most dangerous hurricanes are now far more likely because of climate change, study show.}
\newline
\newline
\emph{Florida Sen. Bill Nelson: Republicans `denying reality' on climate change. ``It's ironic, isn't it?'' Nelson said. ``They accept the hurricane information, but deny the climate information ... Look, some people still think the Earth is flat.''}
\newline
\newline
\emph{Hurricane Harvey's rainfall made three times more likely by global warming, say scientists.}
\newline
\newline
\emph{India heatwave temperatures pass 50 Celsius.}
\newline
\newline
Posts relating to weather are generally about the rise in temperatures in various locations, as well as disasters that occurred as a result of this. 
India is mentioned in relation to extremes in heat and record--breaking temperatures. 
Other locations referred to are the Antarctica seeing high highs in temperatures. 
The topic indicates an overall rise in temperatures across the world and is referred to as a ``crisis.''
This extreme rise in temperatures is also referred to as ``global warming.''

We also see topics at the intersection of politics and climate change. 
\emph{Donald Trump} has been mentioned multiple times over the years mostly in relation to the Paris climate Agreement and talking about climate change from a denial standpoint. 
Some representative documents for this topic are:
\newline
\newline
\emph{Trump begins year--long process to formally exit Paris climate agreement.}
\newline
\newline
\emph{A Late Burst of Climate Denial Extends the Era of Trump Disinformation: A Trump administration official, claiming the imprimatur of the White House, has posted a series of papers questioning the established science of climate change.}
\newline
\newline
\emph{Trump administration scraps clean--water rule aimed at protecting streams, wetlands}
\newline
\newline
\emph{Trump Administration Hardens Its Attack on Climate Science}
\newline
\newline
\emph{Biden Administration to Reinstate Mercury Pollution Rules Weakened Under Trump | The E.P.A. will resume enforcing limits on the release of mercury, a neurotoxin linked to developmental damage in children, from coal-burning power plants.}
\newline
\newline
\emph{White House Moves To Scrap Trump-Era Rewrite Of Key Environmental Law | The Trump rule allowed agencies to ignore climate change when reviewing infrastructure projects and cut the public out of the process.}
\newline
\newline
\emph{Sustainable/Sustainability} has also been mentioned multiple times and in relation to construction, buildings, fashion industry and clothing brands, renewable energy, etc.
Sustainability in relation to fashion and clothing brands is talked about multiple times across the years -- in 2019, 2020, and 2021. 
Posts in this topic are about avoiding fast fashion and opting for more ethical and environment--friendly alternatives. 

The context in which a topic is talked about in a post can help decide the stance the user takes on it. 
For e.g., the topic on climate change news coverage on Fox News with a count of 36, has one representative doc stating: 
\newline
\newline
\emph{``Study: Fox News Climate Coverage Mostly Misleading''} 
\newline
\newline
and another doc stating 
\newline
\newline
\emph{``Fox News climate change coverage is now 28\% accurate, up from 7\%.''}
\newline
\newline
The first one portrays a slight negative standpoint, whereas the latter portrays news coverage in a more positive light. 
Since 2020, there has been a noticeable increase in references to both Covid-19 and climate change as follows:
\newline
\newline
\emph{``How the UK\'s COVID19 lockdown affected average daily electricity demand in March and April. Via Electric Insights by Dr Iain Staffell from Imperial College London for Drax https:\/\/electricinsights.co.uk''} 
\newline
\newline
\emph{``Atmospheric CO2 levels rise sharply despite Covid-19 lockdowns''} 
\newline
\newline
\emph{``Climate crisis to shrink G7 economies twice as much as Covid-19, says research | Climate change''}
\newline
\newline
\emph{``Carbon Dioxide Emissions Rebounded Sharply After Pandemic Dip''}
\newline
\newline
\emph{``CNN: \`Our next thing is going to be for climate change awareness - Climate change is going to be the next COVID thing for CNN~Fear sells.\'''}
\newline
\newline
This indicates that despite the implementation of lockdowns during the Covid-19 pandemic, concerns about climate change persisted, suggesting ongoing environmental consciousness. 
Topics recurring over the years tells us about persistent themes in discussions over time. 
They indicate enduring concerns, ongoing debates, consistent areas of interest, and reflect underlying issues that warrant further investigation or attention. 

\subsection{Named Entities} We extract named entities using the \emph{en\_core\_web\_lg} pipeline of the SpaCy library. 
This has previously been used for research involving social media~\cite{papasavva2020raiders, wang2021multi, balci2023data}.
Examples of entities that can be extracted are locations, times, names, organizations, events, people, etc. 
We are interested in the locations, events and laws for our analysis. 
We use named entity recognition to determine what locations, events and laws are the most talked about in our data.
Table~\ref{tab:locations_NER} shows us the top 15 locations, events and laws and the number of mentions each of these receives in our dataset. 
\begin{table*}[t]
    \begin{tabular}{lclclc}
    \toprule
    Location & \# Frequency & Event & \# Frequency & Law & \# Frequency \\
    \midrule
    Earth & 5,246 & Earth Day & 367 & The Paris Agreement & 357 \\
    Europe & 3,918 & Hurricane Harvey & 199 & The Clean Power Plan & 144 \\
    Arctic & 3,223 & The Great Barrier Reef & 187 & The Inflation Reduction Act & 117 \\
    Antarctica & 1,443 & Hurricane Irma & 111 & The Endangered Species Act & 84 \\
    Africa & 1,327 & Paris Agreement & 111 & Superfund & 69 \\
    Antarctic & 822 & The Green New Deal & 103 & The Paris Climate Agreement & 59 \\
    Asia & 573 & World Environment Day & 87 & Rico & 47 \\
    Atlantic & 573 & Black Friday & 67 & The Clean Water Act & 44 \\
    Pacific & 557 & World War II & 63 & Inflation Reduction Act & 43 \\
    North America & 456 & Hurricane Ida & 53 & Clean Air Act & 34 \\
    Siberia & 368 & Hurricane Ian & 52 & Clean Power Plan & 33 \\
    Arctic Sea & 350 & Olympics & 46 & The Clean Air Act & 33 \\
    West & 349 & The Paris Climate Accord & 45 & Constitution & 33 \\
    Midwest & 262 & Hurricane Maria & 43 & The Montreal Protocol & 30 \\
    Gulf & 246 & Hurricane Dorian & 43 & The Climate Crisis & 27 \\
    \bottomrule
\end{tabular}
\caption{Top 15 locations, events and laws extracted using Named Entity Recognition.}
\label{tab:locations_NER}
\end{table*}
We have mentions of Europe, Africa, Asia and are mostly countries or continents as opposed to smaller and more specific regions and places. 
~\cite{tuitjerSocialMediaPerceived2021} cites Europe to be a popular region to study climate change because of its role in the international fight against climate change. 
We can see that in our dataset, Europe has been referred to 3,918 times. 
Looking back at our results from performing topic modeling, we manually inspect to check for mentions of ``Europe'' and find many topics like:
\newline
\newline
\emph{``European Leaders Agree on Targets to Fight Climate Change - ``Deal!'' Herman Van Rompuy, the president of the European Council, the body that represents European Union leaders, wrote on his Twitter account. ``World's most ambitious, cost-effective'' climate policy agreed on.''}
\newline
\newline
\emph{``Global Warming Forecast Cut on Plans of Biggest Polluters - Climate researchers lowered their forecast for global warming for the first time since 2009 after the greenhouse gas polluters agreed to limit their emissions over the next 15 years, a group of European academics said.''}
\newline
\newline
\emph{Op-Ed - ``It's important that the design of any EU energy strategy is based on real facts and not driven or limited by unsubstantiated emotion or populism.''}
\newline
\newline
\emph{``There is no EU solution to climate change as long as TTIP exists.''}
\newline
\newline
\emph{``The US is banning Russian oil imports, but an embargo that includes European allies would have more impact.''}
\newline
\newline
\emph{``Eco Wave Power's Gibraltar Wave Energy Project Inaugurated and Became Europe's First Grid Connected Wave Energy Array.''}
\newline
\newline
\emph{``China, US and Europe pledge support for global aviation | The proposed new deal on aviation, which aims to cap the carbon pollution of all international flights at 2020 levels will be voluntary between 2021 and 2026 and then mandatory from 2027 for the world's largest emitters.''}
\newline
\newline
\emph{``Why Is There So Much Hype Over Hydrogen? It is reaching new levels of silliness, especially in Europe. The only people who benefit from the hydrogen economy are the oil and petrochemical companies that make the stuff. We may just be at the beginning of a much larger hydrogen hype cycle.''}
\newline
\newline
\emph{``Energy Prices in Europe Hit Records After Wind Stops Blowing -- Natural gas and electricity markets were already surging in Europe when a fresh catalyst emerged: The wind in the stormy North Sea stopped blowing.''}

We also observe that regions like Southern California, Colorado River, North Dakota or NYC have lesser mentions as opposed to the broader regions shown in Table~\ref{tab:locations_NER}. 
These are mentioned 150, 149, 139, and 189 times respectively. 
Colorado River is talked about as:
\newline
\newline
\emph{``The Colorado River By Linn Smith The Colorado River has been over allotted from the beginning, as the Law of the River, a compact made in 1922 between the 7 Colorado River Basin states for the river usage, was made during a time of high precipitation.''}
\newline
\newline
and
\newline
\newline
\emph{``The Colorado River is drying up because of climate change, putting millions at risk of 'severe water shortages.''}
\newline
\newline
Top mentioned events include Earth Day~\cite{EarthDay80:online} which is an annual event to show support for the environment and falls on April 22. 
Similarly, World Environment Day is also an annual event that falls on June 5 and encourages protecting the environment. 
The second most cited event is Hurricane Harvey that happened in 2017. 
Of the 15 events listed, 6 are references to hurricanes. 
The Paris Agreement is the most cited law extracted from our dataset and is also listed as entity ``The Paris Climate Agreement.''
We note that it is also a highly talked about topics as we saw previously from our topic modeling, having generated many topics with representative docs along the lines of:
\newline
\newline
\emph{``While President Donald Trump...is largely dismissive of regulations aimed at tackling climate change, China remains an ardent advocate. China is forecast to achieve its Paris targets on CO2 emissions nearly a decade ahead of schedule.''} 
\newline
\newline
and 
\newline  
\emph{``WHO: Health Benefits Far Outweigh Costs of Meeting Paris Goals. Meeting the goals of the Paris Agreement could save about a million lives a year worldwide by 2050 through reductions in air pollution alone.''}
We also note that many countries like Syria, Nicaragua and India are seen to join the Agreement with positive sentiments and hope.

\section{Discussion and Conclusion}
Our research focuses on studying climate change using Reddit data. 
In this paper, we analyzed climate change--related communities on Reddit over the years 2014 to 2022. 
We put together and analyzed a large dataset that includes posts from 11 subreddits. 
Our analysis provides details about the most talked about topics on these subreddits. 
Topic modeling also revealed interdisciplinary connections within discussions of climate change. 
Some of these topics are water shortage, plant--based food, sustainable fashion, Biden (likely about climate agenda), drought (associated with California), which is related closely to water shortage and rise in temperatures, sea level rise, recycling, republicans' vs Democrats' opinions on climate change, plastics (and in the context of oceans). 
We note topics that talk about climate change from a political point of view and discover topics that have been repeatedly mentioned over multiple years like wildfires, forests, renewable energy, those related to politics, etc. 
We then perform Named Entity Recognition to know more about the locations mentioned in our dataset. 
Of the top 15, ``Earth'' is the most talked about. 
This could be because of posts regarding climate change and concern for the environment across the world as a whole, in addition to its aspects. 
The next on this list is Europe and associated topics also see mentions of wildfires, politics, pollution, and climate action.

In addition to locations, we also extract named entities for events that are referenced as well as laws in place. 
Climate--related events, particularly disasters like hurricanes are among the top mentioned along with environment--related observation days like Earth Day and World Environment Day. 
Following these closely is also the Paris Climate Agreement, be it events or laws in place about climate change. We also saw The Paris Climate Agreement being talked about concerning Donald Trump which has been referenced with many names like The ``Paris Summit'' and ``The Paris Agreement''.

\subsection{Limitations}
Our study comes with certain limitations. 
Studying climate change is something that needs to be done across various disciplines.
While we have posts from 11 subreddits that may touch on these subjects, it still leaves out many more aspects of climate change yet to be researched. 
Three of the eleven subreddits are very large and constitute roughly 70\% of our data. 
We also note that since we rely on embeddings pretrained on subreddits' comments from 2019, we do not account for any new subreddits and comments on them that come about after this period. 
Our analysis focuses only on the posts from said subreddits, and not the comments. 

These limitations provide us with directions for future work on climate change using social media.
This may involve recruiting more subreddits, not limiting future work to just posts but also including comments, and including data after 2022.

\subsection{Acknowledgements}
This material is based upon work supported by the National Science Foundation under Grant No. IIS-2046590.
\bibliographystyle{aaai24}


\appendix

\section{Ethics Checklist}

\begin{enumerate}

    \item For most authors...
    \begin{enumerate}
        \item  Would answering this research question advance science without violating social contracts, such as violating privacy norms, perpetuating unfair profiling, exacerbating the socio-economic divide, or implying disrespect to societies or cultures?
        \answerYes{Yes}
      \item Do your main claims in the abstract and introduction accurately reflect the paper's contributions and scope?
      \answerYes{Yes}
       \item Do you clarify how the proposed methodological approach is appropriate for the claims made? 
       \answerYes{Yes}
       \item Do you clarify what are possible artifacts in the data used, given population-specific distributions?
        \answerNA{NA}
      \item Did you describe the limitations of your work?
      \answerYes{Yes}
      \item Did you discuss any potential negative societal impacts of your work?
        \answerYes{Yes}
          \item Did you discuss any potential misuse of your work?
        \answerNA{NA}
        \item Did you describe steps taken to prevent or mitigate potential negative outcomes of the research, such as data and model documentation, data anonymization, responsible release, access control, and the reproducibility of findings?
        \answerYes{Yes}
      \item Have you read the ethics review guidelines and ensured that your paper conforms to them?
      \answerYes{Yes}
    \end{enumerate}
    
    \item Additionally, if your study involves hypotheses testing...
    \begin{enumerate}
      \item Did you clearly state the assumptions underlying all theoretical results?
        \answerNA{NA}
      \item Have you provided justifications for all theoretical results?
      \answerNA{NA}
      \item Did you discuss competing hypotheses or theories that might challenge or complement your theoretical results?
      \answerNA{NA}
      \item Have you considered alternative mechanisms or explanations that might account for the same outcomes observed in your study?
      \answerNA{NA}
      \item Did you address potential biases or limitations in your theoretical framework?
      \answerNA{NA}
      \item Have you related your theoretical results to the existing literature in social science?
      \answerNA{NA}
      \item Did you discuss the implications of your theoretical results for policy, practice, or further research in the social science domain?
      \answerNA{NA}
    \end{enumerate}
    
    \item Additionally, if you are including theoretical proofs...
    \begin{enumerate}
      \item Did you state the full set of assumptions of all theoretical results?
      \answerNA{NA}
        \item Did you include complete proofs of all theoretical results?
        \answerNA{NA}
    \end{enumerate}
    
    \item Additionally, if you ran machine learning experiments...
    \begin{enumerate}
      \item Did you include the code, data, and instructions needed to reproduce the main experimental results (either in the supplemental material or as a URL)?
        \answerYes{Will be made available upon request.}
      \item Did you specify all the training details (e.g., data splits, hyperparameters, how they were chosen)?
        \answerYes{Yes}
         \item Did you report error bars (e.g., with respect to the random seed after running experiments multiple times)?
        \answerNA{NA}
        \item Did you include the total amount of compute and the type of resources used (e.g., type of GPUs, internal cluster, or cloud provider)?
        \answerYes{Yes}
         \item Do you justify how the proposed evaluation is sufficient and appropriate to the claims made? 
        \answerYes{Yes}
         \item Do you discuss what is ``the cost`` of misclassification and fault (in)tolerance?
         \answerNA{NA}
      
    \end{enumerate}
    
    \item Additionally, if you are using existing assets (e.g., code, data, models) or curating/releasing new assets, \textbf{without compromising anonymity}...
    \begin{enumerate}
      \item If your work uses existing assets, did you cite the creators?
        \answerYes{Yes}
      \item Did you mention the license of the assets?
        \answerNA{NA}
      \item Did you include any new assets in the supplemental material or as a URL?
        \answerNA{NA}
      \item Did you discuss whether and how consent was obtained from people whose data you're using/curating?
        \answerNo{No. We only use publicly available data for our work.}
      \item Did you discuss whether the data you are using/curating contains personally identifiable information or offensive content?
        \answerNA{NA}
    \item If you are curating or releasing new datasets, did you discuss how you intend to make your datasets FAIR (see \citet{fair})?
    \answerNA{NA}
    \item If you are curating or releasing new datasets, did you create a Datasheet for the Dataset (see \citet{gebru2021datasheets})? 
    \answerNA{NA}
    \end{enumerate}
    
    \item Additionally, if you used crowdsourcing or conducted research with human subjects, \textbf{without compromising anonymity}...
    \begin{enumerate}
      \item Did you include the full text of instructions given to participants and screenshots?
      \answerNA{NA}
      \item Did you describe any potential participant risks, with mentions of Institutional Review Board (IRB) approvals?
      \answerNA{NA}
      \item Did you include the estimated hourly wage paid to participants and the total amount spent on participant compensation?
      \answerNA{NA}
       \item Did you discuss how data is stored, shared, and deidentified?
       \answerNA{NA}
    \end{enumerate}
    
\end{enumerate}

\section{Tables}

\begin{table*}[!htb]
    \centering
    \begin{tabular}{lp{7cm}lp{7cm}}
    \toprule
    \textbf{Count} & \textbf{Top 10 Words} & \textbf{Topic Label} & \textbf{Example of a Representative Doc} \\
    \midrule
    698 & fracking, antifracking, frack, chemicals, ban, frackings, boom, texas, drinking, denton & Fracking & ``Fracking and the State'' \\
    508 & ash, duke, coal, carolina, dan, nc, river, spill, north, ponds & Coal Spills & ``Duke Energy Coal Ash Spill in NC, Today.'' \\
    486 & oil, peak, prices, crude, opec, saudi, price, arabia, production, low & Oil Prices & ``US oil production surge to break Saudi Arabia's grip on world energy - Citibank report claims that rising US production over the next decade will erode the role of Saudi Arabia and Opec in supplying the world's largest energy consumer.'' \\
    422 & drought, california, californias, water, lawns, worst, californians, dry, droughts, epic & Drought & ``Study links California drought to global warming'' \\
    412 & wind, turbine, turbines, farm, offshore, farms, airborne, vestas, windfarms, caverns & Wind Turbines & ``Can You Make a Wind Turbine Without Fossil Fuels?'' \\
    381 & nuclear, plants, power, hinkley, nukes, nuke, plant, essential, renaissance, option & Nuclear Plant & ``How nuclear power can stop global warming.'' \\
    327 & panels, solar, panel, solarcity, installation, watt, cheaper, windows, photovoltaic, system & Solar Panels & ``What are the best solar panels for each climate?'' \\
    320 & antarctic, ice, antarctica, sea, sheet, melt, melting, antarcticas, unstoppable, collapse & Ice Melt & ``Antarctic sea ice at record levels despite global warming'' \\
    279 & renewable, renewables, capacity, energy, 100, sources, developing, generation, irena, electrical & Renewables & ``Renewable energy: Back the renewables boom'' \\
    254 & trees, tree, forests, deforestation, forest, tropical, planting, reforestation, ancient, logging & Forests & ``Save Trees. Save Earth'' \\
    \bottomrule
\end{tabular}
\caption{Top 10 topics in 2014.}
\label{tab:topics_in_2014}
\end{table*}

\begin{table*}[t]
    \centering
    \begin{tabular}{lp{7cm}lp{7cm}}
    \toprule
    \textbf{Count} & \textbf{Top 10 Words} & \textbf{Topic Label} & \textbf{Example of a Representative Doc} \\
    \midrule
    1,275 & carbon, co2, emissions, dioxide, greenhouse, tax, capture, atmosphere, gases, pricing & Carbon Emissions & ``Why the global economy is growing, but CO2 emissions aren\'t'' \\
    550 & skeptics, warming, denial, global, skeptic, alarmists, science, alarmism, hoax, conspiracy & Climate Skeptics & ``Climate skeptics\/deniers are now climate change doubters or those who reject mainstream climate science per AP Stylebook entry on global warming'' \\
    532 & fracking, hydraulic, fracturing, pennsylvania, drinking, antifracking, chemicals, ban, epa, sites & Fracking & ``Drinking Water, Fracking, and the EPA'' \\
    458 & pope, francis, encyclical, popes, catholic, vatican, church, catholics, franciss, moral & Catholic Church & ``Pope Francis Says No to Fracking'' \\
    431 & ash, duke, nc, coal, carolina, coalash, energys, ponds, dan, north & Coal & ``Environmental groups challenge North Carolina's coal ash settlement with Duke Energy'' \\
    429 & hottest, warmest, record, year, month, 2014, 2015, was, records, july & Warmest Year & ``Climate change: July was the Earth's hottest month on record -- while 2015 could be the warmest year, scientists say'' \\
    335 & keystone, xl, pipeline, transcanada, veto, pipelines, nebraska, obama, vetoes, senate & Keystone Pipeline & EPA: Keystone XL Pipeline Will Impact Global Warming \\
    319 & obama, obamas, president, legacy, speech, his, barack, union, threat, he & Barack Obama & ``Obama on climate change: `This is a problem now\''' \\
    301 & paris, talks, summit, conference, deal, agreement, bonn, un, negotiators, success &  Paris Summit & ``A Guide to UN climate change talks and the Paris deal'' \\
    293 & prices, oil, crude, price, production, low, crash, glut, decline, collapse & Oil Prices & ``Oil prices fall as industry data shows U.S. crude stocks rising'' \\
    \bottomrule
\end{tabular}
\caption{Top 10 topics in 2015.}
\label{tab:topics_in_2015}
\end{table*}

\begin{table*}[t]
    \centering
    \begin{tabular}{lp{6cm}lp{8cm}}
    \toprule
    \textbf{Count} & \textbf{Top 10 Words} & \textbf{Topic Label} & \textbf{Example of a Representative Doc} \\
    \midrule
    532 & arctic, ice, sea, extent, low, melt, melting, record, lows, minimum & Ice melting & ``Arctic Sea Ice Sets New Record Low For This Time Of Year'' \\
    331 & battery, storage, batteries, lithium, lithiumion, ion, grid, market, grail, holy & Batteries & ``How Distributed Battery Storage Will Surpass Grid-Scale Storage in the US by 2020'' \\
    324 & paris, agreement, deal, ratify, force, ratifies, un, enter, accord, sign & Paris Agreement & ``Paris Climate Agreement - A Good Start'' \\
    323 & oil, prices, peak, barrel, crude, crash, oils, price, iea, cheap & Oil Prices & ``Trump, Clinton and oil prices'' \\
    299 & trump, trumps, donald, president, he, ivanka, denial, elected, presidency, climate & Donald Trump & Trump on Climate Change \\
    246 & floods, louisiana, flooding, flood, rain, rains, deadly, houston, texas, rainfall & Flooding & Scientists See Push From Climate Change in Louisiana Flooding \\
    237 & meat, vegan, eating, beef, vegetarian, eat, veganism, consumption, diets, less & Veganism & ``Eating less meat will reduce Earth's heat (x-post \/r\/vegan)'' \\
    231 & reef, coral, barrier, reefs, great, corals, dead, yacht, dredging, ecosystem & Coral Reefs & ``Over a third of coral is dead in parts of the Great Barrier Reef, scientists say'' \\
    212 & electric, cars, car, vehicles, vehicle, volkswagen, ev, evs, sales, tesla &  Electric Cars & ``Are Electric Cars really Green?'' \\
    209 & green, building, buildings, sustainable, owens, estate, frank, go, sustainability, construction & Sustainability & ``green-renewable-sustainable?'' \\
    \bottomrule
\end{tabular}
\caption{Top 10 topics in 2016.}
\label{tab:topics_in_2016}
\end{table*}

\begin{table*}[t]
    \centering
    \begin{tabular}{lp{6cm}lp{8cm}}
    \toprule
    \textbf{Count} & \textbf{Top 10 Words} & \textbf{Topic Label} & \textbf{Example of a Representative Doc} \\
    \midrule
    696 & pipeline, dakota, keystone, xl, access, pipelines, standing, rock, gallons, protesters & Keystone Pipeline & ``Trump gives new life to Dakota Access, Keystone XL oil pipelines'' \\
    344 & arctic, ice, sea, polar, melting, vortex, frozen, winter, 1979, arctics & Ice Melting & ``Nature, not humans, could be cause of up to half of Arctic sea ice loss, study claims'' \\
    337 & air, pollution, london, quality, deaths, kills, premature, children, linked, malaria & Air Pollution & ``Air Pollution in London'' \\
    291 & wildfires, fires, fire, wildfire, forest, burning, california, californias, smoke, season & Wildfires & ``Scientists See Climate Change in California's Wildfires'' \\
    279 & epa, agency, terminate, staff, looked, protection, staffers, reform, before, abolish & EPA & ``EPA staffer leaves with a bang, blasting agency policies under Trump. `You will continue to undermine your credibility and integrity with EPA staff, and the majority of the public, if you continue to question this basic science of climate change,' Cox wrote.'' \\
    275 & trump, trumps, donald, voters, putin, presidency, ignore, action, policies, closest & Donald Trump & ``Global Anger and Dismay After Trump Slams Brakes on U.S. Climate Action: \"Whoever tries to change into reverse gear is only going to harm themselves when it comes to international competitiveness,\" German environment minister Barbara Hendricks'' \\
    253 & battery, batteries, lithium, storage, lithiumion, breakthrough, ion, flow, solidstate, 2022 & Batteries & ``Are Ice Batteries The Future Of Energy Storage?'' \\
    247 & electric, cars, car, vehicles, evs, vehicle, ev, sales, plugin, tesla & Electric Vehicles & ``Rise of electric vehicles threatens oil industry'' \\
    234 & paris, agreement, accord, withdrawal, decision, exit, leaving, elements, stay, donald & Paris Agreement & ``The US Will Be The Only Country Not In The Paris Agreement. Now What?'' \\
    225 & nuclear, power, reinvent, antinuclear, stranger, option, advanced, lefts, pseg, weapons & Nuclear Power & ``Nuclear energy and climate change'' \\
    \bottomrule
\end{tabular}
\caption{Top 10 topics in 2017.}
\label{tab:topics_in_2017}
\end{table*}

\begin{table*}[t]
    \centering
    \begin{tabular}{lp{6cm}lp{7.5cm}}
    \toprule
    \textbf{Count} & \textbf{Top 10 Words} & \textbf{Topic Label} & \textbf{Example of a Representative Doc} \\
    \midrule
    682 & pipeline, dakota, keystone, xl, access, pipelines, standing, rock, gallons, protesters & Keystone Pipeline & ``Trump orders revival of Keystone XL and Dakota Access pipelines'' \\
    634 & wind, offshore, turbine, farm, turbines, farms, ge, generate, largest, mw & Wind Turbines & ``America's First Offshore Wind Farm'' \\
    494 & paris, agreement, accord, deal, pact, pulling, withdrawal, withdraw, pull, pullout & Paris Agreement & ``The Case for Pulling the U.S. Out of the Paris Climate Accord'' \\
    483 & hydrogen, cell, fuel, hydrogenpowered, h2, cells, toyota, ammonia, fuelcell, hyundai & Hydrogen &``Hydrogen from Ammonia - a better step to hydrogen economy?'' \\
    475 & electric, vehicles, car, cars, ev, charging, evs, vehicle, tesla, volvo & Electric Vehicles & ``EV's and the End of Oil - A look at Electric Cars and the changing Oil Market'' \\
    336 & battery, lithium, batteries, lithiumion, ion, market, storage, usd, liion, rechargeable & Batteries & ``Lithium Ion Battery Market by type and Power Capacity - 2022'' \\
    335 & trump, donald, trumps, ignore, policies, bloomberg, president, mr, climate, leaders & Donald Trump & ``The closest man to Trump is a stealth climate believer. In his new role, Kelly can both reinforce and amplify the views of Secretary of Defense Jim Mattis as he oversees a Pentagon that has long worked to adapt to climate instability both at home and abroad.'' \\
    292 & air, pollution, quality, deaths, london, kills, children, aids, malaria, linked & Air Pollution & ``Air pollution kills...'' \\
    282 & wildfires, fires, fire, wildfire, forest, burning, california, californias, smoke, winds & Wildfires & ``Scientists See Climate Change in California's Wildfires'' \\
    259 & water, drinking, contamination, poll, contaminated, drink, rainwater, tap, conserve, wells & Contaminated Water & ``6 million Americans are drinking contaminated water linked to cancer'' \\
    \bottomrule
\end{tabular}
\caption{Top 10 topics in 2018.}
\label{tab:topics_in_2018}
\end{table*}

\begin{table*}[t]
    \centering
    \begin{tabular}{lp{7cm}lp{7cm}}
    \toprule
    \textbf{Count} & \textbf{Top 10 Words} & \textbf{Topic Label} & \textbf{Example of a Representative Doc} \\
    \midrule
    706 & hydrogen, cell, fuel, cells, h2, hydrogenpowered, toyota, hyundai, fuelcell, powered & Hydrogen & ``Hydrogen Energy Levels'' \\
    641 & renewable, renewables, energy, 100, irena, sources, generation, cost, costs, cheaper & Renewable Energy & ``Can Renewable Energy Power the World?'' \\
    370 & co2, atmosphere, dioxide, ppm, atmospheric, concentration, levels, 415, temperature, earths & Carbon Dioxide & ``CO2 is now past 415 PPM in the atmosphere'' \\
    345 & greta, thunberg, activist, thunbergs, teen, her, summit, she, speech, 16yearold & Greta Thunberg & ``Time Person of the Year: Climate crisis activist Greta Thunberg'' \\
    314 & nuclear, power, answer, opinion, slow, solve, anti, disruption, reasons, academicrepost & Nuclear Power & ``Nuclear Power Can Save the World'' \\
    310 & deal, green, new, principles, oliver, tucker, deals, freedoms, endorsements, radical & Green New Deal & ``What is the Green New Deal?'' \\
    289 & fashion, clothing, clothes, sustainable, wear, fast, brands, cotton, fabric, brand & Sustainable clothing & ``The Fashion Industry and pollution.'' \\
    283 & children, kids, having, parents, child, babies, boys, birthstrike, fewer, childrens & Fewer Children & ``Having children and climate crisis'' \\
    264 & thunberg, greta, thunbergs, her, defence, truant, she, virtue, robbie, biopic &  Greta Thunberg & ``Greta Thunberg'' \\
    238 & antarctic, antarctica, antarcticas, ice, thinning, sheet, thwaites, sea, melting, sheets & Ice Melting & ``The Antarctic is missing 1.25 million km\u00b2 of sea ice'' \\
    \bottomrule
\end{tabular}
\caption{Top 10 topics in 2019.}
\label{tab:topics_in_2019}
\end{table*}

\begin{table*}[t]
    \centering
    \begin{tabular}{lp{6cm}lp{7.4cm}}
    \toprule
    \textbf{Count} & \textbf{Top 10 Words} & \textbf{Topic Label} & \textbf{Example of a Representative Doc} \\
    \midrule
    671 & offshore, wind, turbine, turbines, farm, ge, gamesa, onshore, dogger, farms & Wind Turbines & ``An America offshore wind farm will get the world's largest turbine: The GE Haliade-X'' \\
    452 & renewable, renewables, energy, masters, career, engineer, msc, 100, disproves, jobs & Renewables & ``Why Renewable Energy Is Not Sustainable'' \\
    399 & oil, opec, prices, crude, demand, barrel, price, negative, production, peak & Oil Prices & ``OPEC+ Production Cuts Continue'' \\
    393 & electric, charging, ev, vehicles, vehicle, evs, car, cars, charge, chargers & Electric Vehicles & ``Does Your Climate Action Plan include an EV Electric Vehicle?'' \\
    320 & co2, atmospheric, ppm, dioxide, atmosphere, concentration, co, levels, absorption, earths & Carbon Dioxide Levels& ``Atmospheric Carbon Dioxide Levels Greater than the Past 23 Million-Year Record. CO2 `timeline' revealed no evidence for any fluctuations in CO2 that compares to the dramatic CO2 increase of the present day, which suggests today's greenhouse disruption is unique across recent geologic history.'' \\
    300 & solutions, action, albuquerques, fight, freejah, socially, do, mitigate, change, combat & Climate Action & ``Climate Change Action'' \\
    273 & green, greens, revolution, recovery, investing, greenwash, greening, skerrit, garlands, 5second & Green & ``How to Green the World'' \\
    264 & biden, joe, bidens, presidentelect, his, team, ambitious, he, president, voters & Joe Biden & ``What Does Joe Biden Mean for Our Oceans?'' \\
    261 & fashion, clothing, clothes, fast, sustainable, brands, wardrobe, trashcan, brand, gym & Sustainable Clothing & ``Sustainable clothing and fashion. Is there a market for it?'' \\
    254 & fuels, fossil, subsidies, fuel, defeat, particular, decarbonize, chart, omar, quickly & Fossil Fuel & ``What do we do about fossil fuels?'' \\
    \bottomrule
\end{tabular}
\caption{Top 10 topics in 2020.}
\label{tab:topics_in_2020}
\end{table*}

\begin{table*}[t]
    \centering
    \begin{tabular}{lp{6cm}lp{7cm}}
    \toprule
    Count & Top 10 Keywords & Topic Label &  Example of a Representative Doc \\
    \midrule
    557 & financial, rich, banks, wealthiest, debt, richest, yellen, corporate, finance, pay & Wealthiest and Finances & ``U.S. Warns Climate Poses `Emerging Threat' to Financial System. A Financial Stability Oversight Council report could lead to more regulatory action and disclosure requirements for banks.'' \\
    470 & offshore, wind, farm, floating, farms, turbines, vineyard, approves, ge, administration & Wind Turbines & ``Life at sea by world's largest offshore wind farm'' \\
    380 & nuclear, uranium, atomic, beacon, power, bpn, twh, renaissance, nrc, nuke & Nuclear & ``Nuclear energy'' \\
    380 & covid19, covid, pandemic, coronavirus, pandemics, lockdowns, virus, covid19s, emergence, vaccines & Covid19 Effects & ``Environmental effects of COVID-19 pandemic'' \\
    339 & coal, coalmine, coals, mine, cumbria, coalfired, mines, mining, plants, polands & Coal Mining & ``Cancel All Coal Projects to Have 'Fighting Chance' Against Climate Crisis, Says UN Chief. \"Phasing out coal from the electricity sector is the single most important step to get in line with the 1.5 degree goal.'' \\
    334 & cars, electric, vehicles, car, evs, vehicle, ev, fleet, combustion, electricvehicle & Electric Vehicle & ``Will electric cars stop climate change?'' \\
    329 & air, pollution, deaths, pm25, particulate, quality, premature, exposure, linked, 87m & Air Pollution & ``air pollution and global warming'' \\
    317 & fashion, clothing, clothes, fast, brand, sustainable, brands, wear, sweater, apparel & Sustainable Clothing & ``Fashion and the environment'' \\
    300 & thunberg, greta, blah, activist, she, thunbergs, her, leaders, quote, accuses & Greta Thunberg & ``Climate change: Is Greta Thunberg right about UK carbon emissions?'' \\
    287 & wildfires, wildfire, smoke, fire, fires, bootleg, burning, western, west, forest & Wildfires & ``New Wildfires In Western U.S. At 10-Year High'' \\
    \bottomrule
\end{tabular}
\caption{Top 10 topics in 2021.}
\label{tab:topics_in_2021}
\end{table*}

\begin{table*}[t]
    \centering
    \begin{tabular}{lp{6cm}lp{7cm}}
    \toprule
    Count & Top 10 Keywords & Topic Label & Example of a Representative Doc \\
    \midrule
    876 & wind, offshore, turbine, turbines, farm, floating, farms, blade, onshore, auction & Wind Turbines &``Offshore Wind Farm'' \\
    619 & trees, forests, tree, planting, forest, deforestation, logging, planted, oldgrowth, plant & Forests & ``How much does planting trees really help?'' \\
    535 & sustainability, sustainable, business, corporate, job, businesses, career, grad, development, student & Sustainability & ``I need sustainability help'' \\
    477 & renewable, renewables, energy, energies, switching, 100, transition, electricity, resources, trillions & Renewables & ``What's the state of renewable energy in 2022?'' \\
    303 & amazon, deforestation, rainforest, brazil, brazilian, brazils, lula, indigenous, destruction, forest & Amazon Rainforests & ``Why is Brazil's Amazon rainforest burning?'' \\
    297 & agriculture, farming, regenerative, farmers, agricultural, crop, wheat, crops, yields, vertical & Agriculture & ``Scaling Regenerative Agriculture'' \\
    289 & uae, uaes, emirates, dubai, efforts, arab, 2050, achieve, climatechange, cooperation & Dubai \& Climate Change & ``UAE President Sheikh Mohamed bin Zayed will invest an additional \$50 billion to scale up climate action across the world. Recently, UAE introduce mangrove planting initiative as part of UAE Net Zero by 2050 Strategic Initiative.''\\
    269 & wildfires, wildfire, fire, firefighters, fires, smoke, california, season, californias, mckinney & Wildfires & Wildfire season: How to stop California burning \\
    262 & electric, evs, cars, ev, vehicles, car, vehicle, adoption, sales, toyota & Electric Vehicles & ``Electric cars: should you buy an EV?'' \\
    255 & biden, bidens, emergency, declare, president, joe, executive, declaring, administration, white  & Joe Biden & ``Biden faces pressure to declare a climate emergency'' \\
    \bottomrule
\end{tabular}
\caption{Top 10 topics in 2022.}
\label{tab:topics_in_2022}
\end{table*}

\end{document}